\documentstyle[12pt]{article}
\begin{document}

\begin{titlepage}
\begin{flushright}
{\bf IP-TH-09/97 \\
January 1997 }
\end{flushright}

\vspace{1.5cm}

\begin{center}
{\bf New Solution to Doublet-Triplet Splitting Problem in
SU(N) SUSY GUTs: \\ Towards an Unification of Flavour}\\

\vspace{1.5cm}
{J.L.Chkareuli,~ A.B.Kobakhidze}

\vspace{0.6cm}

{\it Institute of Physics, Georgian Academy of Sciences, 380077
Tbilisi, Georgia $^\star$}
\end {center}

\begin{abstract}
The missing VEV solution to the doublet-triplet splitting problem in
general SU(N) SUSY GUTs is found. Remarkably enough, it requires a
strict equality of numbers of the fundamental colours and flavours in
SU(N) and for the even-order groups (N=2k) gives the predominant
breaking channel $SU(N)\longrightarrow SU(k-1)_c\otimes
SU(k-1)_f\otimes SU(2)_w\otimes U(1)\otimes U(1)^{\prime}$ in the
parameter natural area. The flavour subsymmetry breaking owing to
some generic string inspired extra symmetries of the Higgs
superpotential appears not to affect markedly the basic adjoint
vacuum configuration in the model. Thus the both salient features of
SM just as an interplay between colours and flavours (or families for
due assignment of quarks and leptons) so the doublet structure of the
weak interactions could be properly understood and accommodated in
the framework of the minimal SU(8) model. Among its predictions the
most crucial one belongs to the existence on the TeV scale just three
families of pseudo-Goldstone bosons and their superpartners 3($5+\bar
5$) + SU(5)-singlets which noticeably improve the unification
picture of MSSM.
\end{abstract}

\vspace{6cm}

$\overline{\hspace{5cm}}$

$^\star${\small E-mail: jlc@gps.org.ge~,~ yoko@physics.iberiapac.ge}
\end{titlepage}

\voffset=-2cm

\section{Introduction}

This is a matter of an almost common belief presently that
Supersymmetry plays an essential role in the understanding of the
internal symmetry breaking caused by elementary scalar fields in the
Standard Model and beyond. In the framework of Grand Unified Theories
the scalars of the model should provide through their vacuum
expectation values (VEVs) two gigantically separated breaking scales
$M_{SM}$ and $M_{G}$ as well as guarantee together with a rather
peculiar masses of quarks and leptons a nearly uniform mass spectra
for their superpartners with a high degree of flavour conservation
[1].

However, while SUSY ensures the stability between $M_{SM}$ and
$M_{G}$ against radiative corrections it says nothing about the
very genesis of their hierarchically small ratio $M_{SM}/M_{G}\sim
10^{-14}$ requiring some additional symmetry-based arguments
[2-5]. As to the universal SUSY-breaking terms suggesting
the natural absence  of the flavour-changing neutral currents (FCNC)
it becomes increasingly apparent that together with the large
radiatively induced soft-term non-universality [6] in the
minimal supergravity-based GUTs [7] the supergravity Lagrangian
itself may not have the minimal form as it follows from the direct
higher-order corrections to K\"{a}hler potential [8] as well as a
number of superstring considerations [9]. Presently, among the
other possibilities to alleviate the FCNC problem in SUSY GUTs
some flavour symmetry $G_f$ between quark-lepton (squark-slepton)
families seem to be the most natural framework [10].

Thus contemplating the above SUSY GUT issues (gauge hierarchy and
sfermion democracy) one could hopefully think that they both most
clearly focused on the fundamental Higgs supermultiplet of the SU(N)
GUT might be realized on the one hand in its natural (i.e.
symmetry-conditioned) doublet-triplet mass splitting and on the other
hand in some flavour symmetry in its superpotential couplings with
the basic matter superfields. Further still there could appear in the
GUT symmetry-broken phase some interrelation between the both
phenomena remaining principally down to low energies  in a framework
of supersymmetric unification.

Meanwhile many inventive solutions to the 2/3 problem have been
proposed including the sliding singlet mechanism [2], missing
partner mechanism [3], a special SO(10) based solution [4]
and, possibly, the most physically motivated "Higgs as
pseudo-Goldstone boson" mechanism [5]. Unfortunately, none of
them does concern at all the above second aspect of the
supersymmetric unification which looks necessary to understand why
sfermion spectra exhibit such an unprecedented degeneracy to
adequately suppress FCNC [7-9] without the special
fine tunings just in the sfermion mass matrices
this time.

Although some of those solutions [2-5] are (in principle)
possible one could expect that a full solution to the hierarchy
problem in the SUSY GUTs would naturally appear in models extended
enough to contain the flavour symmetry $G_f$ as well. It seems even
likely that a true 2/3 solution itself might choose the total
starting symmetry of the GUT including some additional symmetry for a
flavour in the model.

Following such a motivation [11] we consider below general SU(N)
SUSY GUTs.  We find that under certain and quite acceptable
circumstances there can be developed some new missing VEV type
mechanism for 2/3 splitting which definitely favours just
the SU(8) GUT among the other SU(N) theories.

\section{Missing VEV solutions in SU(N) GUTs}

The missing VEV ansatze as it was formulated by Dimopoulos and
Wilczek a long ago [3] is that a heavy adjoint scalar $\Sigma
^i_j$ ($i,j=1...N$) of SU(N) might not develop a VEV in the weak
SU(2) direction and through its coupling with fundamental chiral
pairs $H_i$ and $\overline{H^i}$ (containing the ordinary Higgs
doublets which break an electroweak symmetry and give masses to up
and down quarks, respectively, in a final stage of symmetry breaking)
could hierarchically split their masses in the desired 2/3 way.
While they found some special realization of the missing VEV
mechanism in SO(10) model [3] the situation in SU(N) theories
looks much hopeless.  The main obstacle is happened to be a presence
of a cubic term $\Sigma ^3$ in the general Higgs superpotential W
leading to the impracticable trace condition $Tr\Sigma
^2=0$\footnote{In a standard SU(5) model the missing VEV solution
$<\Sigma>=\sigma diag(1~1~1~0~0~0)$ is trivially impossible even  by
an ordinary trace condition $Tr\Sigma=0$.} for the missing VEV vacuum
configuration unless there occurs the special fine-tuned cancellation
between $Tr\Sigma ^2$ and driving terms stemming from the other
parts of the W [3].

So, the only way to a natural missing VEV solution in SU(N) theories
seems to exclude the cubic term $\Sigma ^3$ from the superpotential W
imposing some extra reflection symmetry on the adjoint supermultiplet
$\Sigma$
\begin{equation}
\Sigma \longrightarrow -\Sigma
\label{1}
\end{equation}
While an elimination of the $\Sigma ^3$ term itself leads usually to
the trivial unbroken symmetry case an inclusion of the higher
even-order $\Sigma$ terms (supposedly induced by gravity) in the
effective superpotential or calling into play another adjoint scalar
$\Omega$ as well (even under all the renormalizable couplings in W)
can, as we show below, drastically change a situation. Although the
both approaches are proved to be closely related [12] we consider
rather from force of habit the two-adjoint case first.

The general renormalizable two-adjoint Higgs superpotential of
$\Sigma$ and $\Omega$ satisfying the reflection symmetry (1) is
\def\theequation{2a}
\begin{equation}
W_{adj}=\frac{M_1}{2}\Sigma ^2+\frac{M_2}{2}\Omega
^2+\frac{h}{2}\Sigma ^2\Omega+\frac{\lambda}{3}\Omega ^3
\label{2a}
\end{equation}
The total superpotential W apart from the adjoints
$\Sigma$ and $\Omega$  as well as the ordinary Higgs-doublet
containing chiral superfields H and $\overline H$ includes also $N-5$
fundamental chiral pairs $(\varphi,~~\overline{\varphi})^{(r)}$
$(r=1...N-5$) which break SU(N) to SU(5) at some high mass scale
(possibly even at $M_{GUT}$) \def\theequation{2b} \begin{equation}
W=W_{adj}+W_{H}+W_{\varphi}
\label{2b}
\end{equation}
We do not consider for a moment the
$W_{\varphi}$ part of the superpotential assuming that some extra
symmetry (see Sec.3) makes it possible to ignore its influence on the
formation of the basic vacuum configurations in the model.

One can find from the vanishing F-terms of the adjoints $\Sigma$
and $\Omega$ in $W_{adj}$ (in matrix notation for their VEVs)
\def\theequation{3a}
\begin{equation}
M_1\Sigma +h[\Sigma
\Omega -\frac{1}{N}Tr(\Sigma \Omega)]=0,
\label{3a}
\end{equation}
\def\theequation{3b}
\begin{equation}
M_2\Omega +\frac{h}{2}[\Sigma
^2 -\frac{1}{N}Tr\Sigma ^2]+\lambda[\Omega
^2-\frac{1}{N}Tr\Omega ^2]=0
\label{3b}
\end{equation}
that all the basic $\Sigma$ and $\Omega$ vacuum configurations
related with superpotential $W_{adj}$ include the following four
classes: \\
(i)~The trivial (symmetry-unbroken) case
\def\theequation{4}
\begin{equation}
\Sigma =0,~~~~\Omega =0
\label{4}
\end{equation}
(ii)~The single-adjoint like cases (as if the adjoint
$\Omega$ was alone in $W_{adj}$)
\def\theequation{5a}
\begin{equation} \Sigma =0, \label{5a}
\end{equation}
\def\theequation{5b}
\begin{equation}
\Omega =\omega _SR^{(m,N-m)},~~\omega
_S=-\frac{M_2}{\lambda}\frac{N-m}{N-2m}
\label{5b}
\end{equation}
where $R^{(m,N-m)}$ is a familiar diagonal matrix of type
\def\theequation{5c}
\begin{eqnarray}
~\hspace{4.cm}m\hspace{2.3cm}N-m\hspace{1.2cm}
\nonumber \\
R^{(m,N-m)}=diag(\overbrace{~1~...~1~}~,
~\overbrace{-\frac{m}{N-m}...-\frac{m}{N-m}})
\label{5c}
\end{eqnarray}
(iii)~The "parallel" VEVs ($\Sigma \propto \Omega$) \\
\def\theequation{6a}
\begin{equation}
\Sigma=\omega_{P} R^{(m,N-m)},~~~\sigma
_P=2^{1/2}\frac{M_1}{h}\frac{N-m}{N-2m}\biggl(\frac{M_2}{M_1}
-\frac{\lambda}{h}\biggl)^{1/2}~,
\label{6a}
\end{equation}
\def\teequation{6b}
\begin{equation}
\Omega=\omega _{P} R^{(m,N-m)},~~~\omega
_P=-\frac{M_1}{h}\frac{N-m}{N-2m}
\label{6b}
\end{equation}
(iv)~The "orthogonal" VEVs ($Tr(\Sigma \Omega)$ and thus
$Tr(\Sigma \Omega^2)=Tr\Sigma ^3=0$ as it follows from basic Eqs.(3))
\def\theequation{7a}
\begin{equation}
\Sigma=\sigma _{O}R^{(\frac{m}{2},\frac{m}{2},N-m)},~~~\sigma
_O=2^{1/2}\frac{M_1}{h}\frac{N}{N-m}\biggl(\frac{M_2}{M_1}
-\frac{\lambda}{h}\frac{N-2m}{N-m}\biggl)^{1/2}~,
\label{7a}
\end{equation}
\def\theequation{7b}
\begin{equation}
\Omega=\omega _O~R^{(m,N-m)},~~~\omega
_O=-\frac{M_1}{h}
\label{7b}
\end{equation}
where a new diagonal matrix $R^{(\frac{m}{2},
\frac{m}{2},N-m)}$
\def\theequation{7c}
\begin{eqnarray}
~\hspace{5.5cm} m/2\hspace{1.1cm} m/2\hspace{1.2cm} N-m
\nonumber \\
R^{(\frac{m}{2},\frac{m}{2},N-m)}=(\overbrace{~1~...~1~}~,
\overbrace{~-1~...~-1~}~, \overbrace{~0~...~0~})
\label{7c}
\end{eqnarray}
is just  orthogonal to matrix $R^{(m,N-m)}$ of type (5c)
with a proper value of $m$. The group decomposition numbers $m$
($m=1...N$) are different in general in all the  above classes.

So, we can conclude that while an "ordinary" adjoint $\Omega$ having a
cubic term in $W_{adj}$ (3) develops in all non-trivial cases
(ii - iv) only a "standard" VEV of type (5c) which breaks the
starting symmetry SU(N) to $SU(m)\otimes SU(N-m)\otimes U(I)_{S}$ the
second adjoint $\Sigma$ can have also a new orthogonal solution of
type (7c) with $SU(N)$ breaking along the channel $SU(m/2)\otimes
SU(m/2)\otimes SU(N-m)\otimes U(I)_O\otimes U(I)_S$. This case
corresponds to the missing VEV solutions just we are looking for if
one identifies an $SU(N-m)$ subgroup of $SU(N)$ with the weak symmetry
group while two other $SU(m/2)$ groups should be identified therewith
the groups of the fundamental colours and flavours, respectively,
\def\theequation{8}
\begin{eqnarray}
N=n_c+n_f+n_w~,~~~~~~~~~~~~~~~~~~~~~~~           \\
n_c=m/2,~~n_f=m/2,~~n_w=N-m \nonumber
\end{eqnarray}
If so, we are driving at general conclusion that a missing
VEV solution in $SU(N)$ theories appears only when the numbers of
colours and flavours\footnote{Or families if one takes a proper
assignment for quarks and leptons under the flavour group $SU(m/2)_f$
(see Sec. 4)} are happened to be equal.

Meanwhile there are other solutions (4-6) which together with the
missing VEV one give a four-class vacuum degeneracy related with the
superpotential $W_{adj}$ (2). We show now that a supergravity-induced
lifting the vacuum degeneracy naturally singles out among the other
ones just the missing VEV configuration (7) for a favourable
parameter space in superpotential (2) and for the actually observed
subgroup content of SU(N).

Indeed, the inclusion of supergravity modifies the form of the
effective potential at low energies [1] so that we have for a
potential at the minimum now (to the lowest order in $k=M^{-1}$, $M$
is Planck mass)
\def\theequation{9}
\begin{equation}
V_{adj}\simeq -3k^2|W_{adj}|^2
\label{9}
\end{equation}
which gives different values for the above vacuum
configurations\footnote{In general a lifting the vacuum
degeneracy will induce a negative cosmological constant which,
however, can be cancelled by a slight redefinition of the hidden part
of the superpotential. We take that the absolute minimum of the
potential $V_{adj}$ which singles out the true vacuum
configuration for the SU(N) symmetry breaking can always be
arranged at $E=0$ (while the other minima lie higher) by a
proper enlargement of the adjoint scalar sector [13].} inasmuch as
there are, respectively,
\def\theequation{4$^{\prime}$}
\begin{equation}
W^{O}_{adj}=0~,
\end{equation}
\def\theequation{5$^{\prime}$}
\begin{equation}
W^{S}_{adj}=\frac{4}{27}\alpha N\frac{a+1}{a^2}r^3~,
\end{equation}
\def\theequation{6$^{\prime}$}
\begin{equation}
W^{P}_{adj}=\alpha N\frac{a+1}{a^2}(r-1)~,
\end{equation}
\def\theequation{7$^{\prime}$}
\begin{equation}
W^{O}_{adj}=\alpha N(a+1)(r+a)
\end{equation}
Here new parameters $\alpha$, $r$ and $a$ expressed (in
terms of the adjoint masses and coupling constants as well as group
parameters $N$ and $m$ are  just
\def\theequation{10}
\begin{equation}
\alpha =\frac{\lambda}{3} \frac{M_1^3}{h^3}~,
~~r=\frac{3h}{2\lambda}\frac{M_2}{M_1}~, ~~a=-\frac{N-2m}{N-m}
\end{equation}

An inspection of Eqs.($5^\prime$, $6^\prime$, 10) shows that the
maximal values of $W^{S}_{adj}$ and $W^{P}_{adj}$ correspond to the
minimal possible value of $a$  which is to say that $|N-2m|=1$ or 2~
($N-2m=0$ gives no symmetry breaking in the S and P cases, see the
basic Eqs.(3)) depending on whether one starts with the odd
$(N=2k-1)$ or even $(N=2k)$ order SU(N) group.  Meanwhile for the
missing VEV solutions (7) a maximal value of $W^{O}_{adj}$
($7^\prime$,10) corresponds to the maximal possible value of $a$ or a
minimal one of $n_w= N-m$ (see Eq. (8)). So far as a group
decomposition number $m$ is, by definition, the even number in the
missing VEV case (7) the latter means that all the odd-order $SU(N)$
groups single out the unrealistic vacuum configurations with $n_w=1$
(and thus should be excluded) while the even-order ones drive exactly
at $n_w=2$.  Finally, a prerequisite to the formation of the global
minimum of the potential $V_{adj}$ (9) for the dominant $n_w=2$
missing VEV solution versus those from the alternative classes (4-6)
leads to the natural restriction on the dynamical parameter $r$
\def\theequation{11}
\begin{equation}
-3\biggl(\frac{N-2}{N+2}\biggl)^{1/3}~<~r~<~3\biggl
(\frac{N-2}{N+2}\biggl)^{1/3}
\end{equation}
as it follows directly from a comparison of $W^O_{adj}$ with
$W^{S}_{adj}$ and $W^{P}_{adj}$ in O(1/N) approximation.

In addition to the basic (one parametrical) vacuum configurations
(5-7) some their non-trivial superpositions can also appear in the
two-adjoint case considered. However, all of them are found [12] not
to be essential in the interval (11) where the key missing VEV
configuration principally evolves.

So, one can conclude that the realistic missing VEV solution
naturally appears in all the even-order $SU(N)$
symmetry-contained theories globally dominating in the most favourable
parameter area (11) where all masses and coupling constants can have
the same order values. Remarkably, the solution gives an insight
into why the numbers of the fundamental colours and flavours are
happened to be equal and provides an explanation for the weak  $SU(2)$
symmetry structure coming safely from a grand unified scale down to
low energies.

In essence, there is only one parameter left for a final
specialization of theory -- the number of colours. It stands to reason
that $n_c=3$ and we come to the unique $SU(8)$ symmetry case with the
missing VEV breaking channel
\def\theequation{12}
\begin{equation}
SU(8)\longrightarrow SU(3)_c\otimes SU(3)_f\otimes SU(2)_w\otimes
U(I)_O\otimes U(I)_S
\end{equation}
which certainly dominates over the other possible ones in
the natural $r$ parameter area (11).


\section{Flavour symmetry breaking}

Let us consider now the other parts of the total Higgs superpotential
$W(2b)$. There $W_H$ is in fact the only reflection-invariant
coupling of the adjoint $\Sigma$ with a pair of the ordinary
Higgs-doublet containing fundamental chiral superfields $H$ and
$\bar{H}$
\def\theequation{13}
\begin{equation}
W_{H}=f\bar{H}\Sigma H, ~~~~~~(\Sigma \longrightarrow
-\Sigma,~~~~~ \bar{H}H\longrightarrow -\bar{H}H)
 \end{equation}
having the zero VEVs $H=\bar{H}=0$ during the first stage of the
 symmetry  breaking. Thereupon $W_H$ turns to the mass term of $H$
and $\bar{H}$ fields depending on the basic vacuum configuration
(4-7) in the model. The point is the $\Sigma$ missing VEV
configuration giving generally heavy masses (of the order $M_G$) to
them leaves their weak components strictly massless. Thus there
certainly is a natural doublet-triplet splitting in the model
although we drive at the vanishing $\mu$-term on this stage. One can
argue that some $\mu$-term always appears through the radiative
corrections [1,7] or a non-minimal choice of K\"{a}hler potential [8]
or the high-order terms induced by gravity (see below) in the flavour
part of the superpotential $W_{\varphi}$ we are coming to now.

The flavour symmetry breaking which is also assumed to happen  on the
GUT scale $M_{G}$ (not to spoil the standard supersymmetric grand
unification picture) looks in the above favoured $SU(8)$ case (12) as
\def\theequation{14}
\begin{equation}
 SU(3)_f\otimes U(I)_O\otimes U(I)_S \longrightarrow
 U(1)_Y
 \end{equation}
where $U(I)_{S,O}$ are hypercharges given by the matrices
(5c) and (7c), respectively, while $U(1)_Y$ is an ordinary $SU(5)$ one
[1].

A question arises: how the missing VEV solution (7) can survive such
a high-scale symmetry breaking (14) so as to be subjected at most
to the weak scale order shift? The simplest way  could be if there
appeared some generic discrete symmetry $Z_k$ which forbade the mixing
between  two sectors $W_{adj}+W_{H}$ (I) and $W_{\varphi}$ (II) in
the Higgs superpotential\footnote{Such a way was intensively discussed
in the context of the "Higgs as PGB" mechanism [5] and the
several particular (predominantly non-renormalizable) models were
constructed.  Another scenario suggested recently in the same context
[5] uses the anomalous $U(1)_A$ symmetry which can naturally get
untie the sectors (I) and (II) and induce the high-scale family
symmetry breaking through the Fyet-Illiopoulos D-term (see the last
paper in Ref.[5]).  It stands to reason that both ways are fully
suited for our case as well although the second one looks less
instructive when the flavour symmetry is high as in (14).}. Clearly,
in a general SU(N) GUT the accidental global symmetry
$SU(N)_I\otimes SU(N)_{II}$ (and possibly a few $U(1)^\prime$s) being
a result of a Z-symmetry must appear and PGB$^\prime$s are produced
after the global symmetry breaks. On the other hand Z-symmetries tend
to strongly constrain a form of superpotential $W_{\varphi}$ itself
so that contrary to the supersymmetric adjoint breaking (5-7) the
flavour symmetry of SU(N) can break triggered just by the soft SUSY
breaking only. As it takes place, Z-symmetries are happened to
set up some hierarchy of the flavour breaking scale $M_f$  with
respect to the Planck scale M and the soft SUSY breaking scale m.

We briefly run through a possible scenario of the flavour symmetry
breaking which could  naturally appear in the renormalizable SU(N)
theories.

Let there are some discrete symmetries $Z_3^r$ for each pair $r$
($r=1 ... N-5$) of the flavour chiral superfields $(\bar{\varphi},
\varphi) ^{(r)}$ in the superpotential $W_{\varphi}$
\def\theequation{15}
\begin{equation}
 \bar{\varphi}_r\varphi_r \longrightarrow Z^r_3~
 \bar{\varphi}_r\varphi_r~~~~(Z^r_3 :~\omega _r=e^{i2\pi/3})
 \end{equation}
These $Z_3^{r}$ should actually be gauge type discrete
symmetries stable under gravitational corrections [14].They are
assumed to be inherited from Superstrings  so that the   discrete
anomalies of $\varphi^\prime$s could always be cancelled by adding extra
gauge singlets Y (string modes) transforming non-trivially under
$Z_3$-symmetries.  So, their
common superpotential $W_{\varphi}$ could have the following general
form
\def\theequation{16}
\begin{equation}
W_{\varphi}= a_{rs} Y_{rs} \bar{\varphi}_r\varphi_s
+\frac{\delta_{rs}}{3}[b_{rs} Y_{rs}^3 + c_{rs}\bar{Y}_{rs}^3 +
3 M \bar {Y}_{rs}Y_{rs}]
\end{equation}
where $a_{rs}$, $b_{rs}$ and $c_{rs}$ are matrices of the
coupling constants (summation is meant over all r, s values, $\delta
_{rr}=1$). The massless non-diagonal (on the $\varphi$ species) and
massive diagonal $Y$-fields in $W_{\varphi}$ (16) could be considered
as the basic superstring gauge singlet modes having the zero and
Planck scale $M$ order masses, respectively.

 One can see that the total accidental global symmetry of the
 Higgs-superpotential $W$ (2,13,16) caused by discrete symmetries
(1,15) apart from the above mentioned $SU(N)_I\otimes SU(N)_{II}$
 symmetry includes also the $U(1)^\prime$s concerning all the scalar
 species
\def\theequation{17}
\begin{equation}
 U(1)_H\otimes \prod\limits^{N-5}_r U(1)_r
\end{equation}
Whereas the diagonal $Y$ fields are neutral under the $U(1)$
 symmetries (17) the non-diagonal ones carry those quantum numbers as
 well
\def\theequation{18}
\begin{equation}
Y_{rs} \longrightarrow Z_3^r e^{i(Q_r-Q_s)\alpha}Y_{rs} ~~~~~(Z_{3}^r
: \bar{\omega}_r= e^{-i2\pi/3})
\end{equation}
($Q_{rs}$ are hypercharges corresponding to $U(1)_{rs}$).
Just this symmetry (18) prevents them from having any self-interacting
terms in the invariant superpotential $W_{\varphi}$(16).

 Now the standard analysis of the F-terms of the $Y$-fields reveals
 the specific (inspired by $Z_3^r$ symmetries) hierarchical
 relations between VEVs of scalars in the supersymmetric limit
\def\theequation{19a}
\begin{equation}
\frac{\bar{Y}_{rs}}{M}=\delta_{rs}
\biggl(-\frac{1}{c_{rr}}\biggl)^{1/2}
\biggl(\frac{Y_{rr}}{M}\biggl)^{1/2}
\end{equation}
\def\theequation{19b}
\begin{equation}
\frac{\bar{\varphi}_{r}\varphi_s}{M^2}~=~-\frac{\delta_{rs}}{a_{rr}}
\biggl[\biggl(-\frac{1}{c_{rr}}\biggl)^{1/2}
\biggl(\frac{Y_{rr}}{M}\biggl)^{1/2}+O\biggl(\frac{Y_{rr}^2}{M^2}\biggl)\biggl]
\end{equation}
while the vanishing F-terms of the flavour superfields
$\varphi^{(r)}$ make all the $Y_{rr}$ to have zero VEVs in that
limit.
However, as one can explicitly show the soft breaking terms in the
total scalar potential
\footnote {D-terms are assumed to be vanished not to have a large SUSY
breaking in the visible sector.}
\def\theequation{20}
\begin{equation}
V=\sum \limits_k \biggl|\frac{W}{\phi_k}\biggl|^2~+~A m W^{(3)}~+~B m
W^{(2)}~+~m^2 \sum\limits_k |\phi_k|^2 \end{equation}
(where $\phi_k$ denote all the fields included in the total
superpotential $W$ with its bilinear $W^{(2)}$ and
trilinear $W^{(3)}$ parts, respectively) will inevitably shift the
VEV of $Y_{rr}$ from zero to \def\theequation{21} \begin{equation}
Y_{rr}~=~\frac{xm}{a_{rr}}
\end{equation}
(x is a dimensionless parameter depending on the soft breaking
parameters $A$ and $B$ in the potential V) and lead to the flavour
symmetry breaking according to the Eq.(19b).  For  $m\sim~10^3$ GeV
and coupling constants (primarily $a_{rr}$) of order 0.1 in the
superpotential $W_{\varphi}$ one can have for a flavour
scale\footnote{Also, any higher flavour symmetry breaking scale $M_f$
could be generated if one took the higher discrete symmetry $Z_n$
(and the corresponding number of the massive gauge singlets "softly"
converting the $(\bar{\varphi} \varphi)^{(r)}$ pairs into vacuum).}
$M_f\sim M_G$ as it is wanted for a single point grand
unification.

Another remarkable feature of the pattern (19b) is the flavour scalars
$\varphi^{(r)}$ develop generally their VEVs in all the $N - 5$
orthogonal direction so that the flavour part of the starting SU(N)
gauge symmetry turns out to be completely broken and one comes safely
just to the SM.

Whereas there does not appear any appreciable influence on the
missing VEV solution (7) due to the above flavour symmetry breaking
mechanism some gravitational corrections which mix the sectors (I)
and (II) are generally expected. The largest mixing term allowed by
$Z_3^r$ symmetries is
\def\theequation{22}
\begin{equation}
A_r\bar{\varphi}_r \Omega \varphi_r
\biggl[\frac{\bar{\varphi}_r {\varphi_r}}{M^2}\biggl]
^2~~~(A_r=0.01\div 1)
\end{equation}
which according to the basic Eqs. (3 a,b) will lead to the
shift in the VEVs of the adjoint $\Sigma$ and $\Omega$ (and
subsequently to the $\mu$-term for Higgs doublets in $H$ and
$\bar{H}$) just of the order
\def\theequation{23}
\begin{equation}
A_rM_G\biggl(\frac{M_G}{M}\biggl)^4~\sim~ 10^2~ \div ~10^4~GeV
\end{equation}
The same welcome order is expected for masses of the
PGB$^\prime$s inasmuch as the mixing term (22) breaks explicitly the
global $SU(N)_I \otimes SU(N)_{II}$ to $SU(N)$. So, one can see that
the gravitational corrections make the model even more realistic than
it could do the ordinary $SU(N)$ gauge ones if they were alone.

\section{Particle spectra and unification}

The time is right to discuss now the particle spectra in the model
concentrating mainly on its favoured minimal $SU(8)$ version (12).
 Having considered the basic matter superfields (quarks and leptons
 and their superpartners) the question of whether the above flavour
 symmetry $SU(3)_f$ is yet their family symmetry naturally arises.
 Needless to say that among many other possibilities the special
 assignment treating the families as the fundamental triplet of
 $SU(3)_f$ comes first.

In such a case the anomaly-free set of $SU(8)$ antisymmetric
multiplets
\def\theequation{24a}
\begin{equation} 6\cdot 8^A +
28^{[AB]} + 2\cdot 56_{[ABC]} + 70_{[ABCD]}
\end{equation}
is singled out if we require that after flavour symmetry
breaking (14) only three massless families ordinary quarks and
leptons (and their superpartners) are left as a chiral triplets of
$SU(3)_f$ stemming from multiplets
\def\theequation{24b}
\begin{equation}
\overline {28} = (\bar5,\bar3)+ ...,~~~~70=(10,\bar3)+ ... ,
\end{equation}
while the rest $SU(5)\otimes SU(3)_f$ fragments\footnote{One  uses
(for convenience) here and below the reps of $SU(5)$ for the standard
particle assignment only without any reference to it as an
intermediate symmetry.} in them as well in other multiplets (24a)
acquire heavy masses of order $M_f\sim M_G$.

So, one drives again at the known chiral $SU(3)_f$ family symmetry
case [15] following this time from the special multiplet arrangement
(24a)\footnote{Remarkably, the multiplets (24a) follow from the
unique ("each multiplet - one time") set of the $SU(11)$ multiplets
[16] after symmetry breaking $SU(11) \longrightarrow SU(8)$ and the
exclusion of all the conjugated (under $SU(8)$)) multiplets but the
self-conjugated one $70_{[ABCD]}$.} considered before one of us [17]
as a possible base for the family-unifying $SU(8)$ GUT. There the
universal see-saw mechanism inducing the non-trivial fermion
mass-matrices (with many texture ansatzes available) is proved
to be developed so that the observed pattern of quark (and
lepton) masses and mixings can appear after the electroweak
$SU(2)\otimes U(1)_Y$ symmetry breaks [16].

Meanwhile due to the
absence of the direct trilinear couplings of the fermion multiplets
(24b) containing quarks and leptons with Higgs octets H and $\bar H$
there do not appear in general any "vertical" mass relation between
quarks and leptons inside of a family like as known $b-\tau$
unification in the simplest SU(5) model [1]. Such a unification if
appeared in the present SU(8) model would lead to the unacceptable
heavy b-quark at low energies (see below).

Another block, or it would be better to say a superblock of the
low-energy particle spectrum is the PGB$^\prime$s and their
superpartners$^{7}$
\def\theequation{25}
\begin{equation}
3(5+\bar5)~+~SU(5)-singlets
\end{equation}
appearing as a result of breaking of an accidental
$SU(8)_I\otimes SU(8)_{II}$ symmetry in a course of the spontaneous
breakdown of the starting local $SU(8)$ symmetry to SM (12, 14) and
acquiring a weak scale order masses due to the soft SUSY breaking and
subsequent radiative corrections or directly through the
gravitational corrections of type (22). Normally they are the proper
superpositions of the $SU(5)\otimes SU(3)_f$ fragments $(5,\bar3) +
(\bar5,3)$ in the adjoints $\Sigma$ and $\Omega$  of $SU(8)$ and
$(5,1)_r+ (\bar5,1)_r$ $(r=1,2,3)$ in its fundamental flavour scalars
$\varphi_r$ (see above).

So, at a low-energy scale one necessarily  has in addition to three
standard families of quarks and leptons (and squarks and sleptons)
just three families of PGB$^\prime$s and their superpartners which,
while beyond the one-loop approximation in the renormalization group
equations (RGEs) will modify the running of gauge and Yukawa
couplings in the model.

We found that the MSSM predictions for $\alpha _S(M_Z)$ and $M_G$
changed as the PGB supermultiplets were included in the RGEs mainly
to the top-Yukawa coupling corrections to the evolution of the
standard gauge couplings $\alpha _1$, $\alpha _2$ and $\alpha
_S$\footnote{Without considering those corrections the MSSM
predictions were found [19] to be practically stable under various
extensions of MSSM with extra complete SU(5) multiplets at the weak
scale.}. These corrections having been calculated in the overall
two-loop approximation [18] are happened to be quite noticeable for
the starting large values of the top-Yukawa coupling on the GUT scale
$Y_t(M_G)\geq 0.1$ ($Y_t=h_t^2/4\pi$) evolving rapidly towards its
infrared fixed point.  \begin{table}[t] \caption{\small The SU(8)
model vs MSSM (in square brackets) predictions for $\alpha _S(M_Z)$,
$tan\beta$, $\alpha _G$ and $M_G$ (the input parameters are presented
in Eqs. (26-27)).} \begin{center}
\begin{tabular}{|c|c|c|c|c|c|}\hline
&&&&&\\
$T_{SUSY}$ & $Y_t(M_G)$ & $\alpha_S(M_Z)$ & $tan\beta$ &
$\alpha _G$ & $M_G/10^{16}$ GeV \\ &&&&&\\ \hline
&&&&&\\ & 0.1 & 0.121 [0.126] & 1.08 [1.55] & 0.143 [0.042] & 5.17
[3.03] \\ $M_Z$ & & & & & \\ & 0.3 & 0.120 [0.124] & 1.07 [1.44] &
0.138 [0.042] & 5.05 [2.91] \\ &&&&&\\ \hline &&&&& \\ & 0.1 & 0.120
 [0.123] & 1.15 [1.69] & 0.130 [0.041] & 4.02 [2.5] \\
$m_t^{pole}=175$ GeV & & & & & \\ & 0.3 & 0.119 [0.122] & 1.14 [1.54]
 & 0.125 [0.041] & 3.92 [2.4] \\ &&&&& \\ \hline \end{tabular}
 \end{center} \end{table}

Our results summarized in Table 1. We used as inputs on the one hand
the electroweak scale parameters (the World average ($\overline
{MS}$) central values [20])
\def\theequation{26}
\begin{equation}
\alpha _{EM}(M_Z)=1/127.9,~~ sin^2\theta _W(M_Z)=0.2313,~~
m_t^{pole}=175~ {\rm GeV}
\end{equation}
and on the other hand the GUT scale ones -- some tolerable top-Yukawa
coupling and PGB mass values on $M_G$
\def\theequation{27}
\begin{equation}
Y_t(M_G)=0.1~ {\rm and}~ 0.03,~~~ \tilde {\mu}(M_G)=600~ {\rm GeV}
\end{equation}
to determine $\alpha _S(M_Z)$ and $tan\beta$, as well as SU(8)
unifying parameters $\alpha _G$ and $M_G$ (in $\overline {DR}$
sheme).

Performing the running of the gauge and top-Yukawa couplings from the
GUT scale down to $M_Z$ (by considering two-loop $\beta$-functions
[18] up-dated for the SU(8) case with PGB supermultiplets) the
threshold corrections related with the doublet-triplet splittings in
the PGB states (25) due to one-loop RGE evolution [19] as well as
overall one-loop supersymmetric threshold corrections associated with
the decoupling of the supersymmetric particles at some effective
(lumped) scale $T_{SUSY}$ have also been included. We took for
$T_{SUSY}$ two the relatively low values (which in one way or another
are singled out) $T_{SUSY}=M_Z$ and $T_{SUSY}=m_t^{pole}$ to keep
sparticle masses typically in a few hundred GeV region.

As one can see from Table 1 the $\alpha _S(M_Z)$ values  are markedly
closer to the most stable World average value [20] $\alpha
_S(M_Z)=0.118\pm 0.003$  in the SU(8) model than in MSSM and, in
substance, there is a good agreement with data for $Y_t(M_G)\geq
0.1$. Thus gauge coupling unification seems to favour the SU(8) model
over MSSM though a total uncertainty (experimental and theoretical)
in $\alpha _S(M_Z)$ is still left too large to see it distinctively.

Meanwhile the $b-\tau$ unification as it directly stems from our
calculations (see also [19]) proves to be very sensitive to the
presence of new PGB states (25) in our model and actually
breaks unless $m_b(m_b)\geq 5.5$ GeV what seems to be excluded by
experiment [20]. Fortunately, as it was mentioned above the
model does not predict $b-\tau$ unification in general and thus it
can not be considered as its critical test.

Another significant outcome of the SU(8) model turns out to be very
low (closed to one) values of $tan\beta$ if one takes the $Y_t$
fixed-point solution (see Table 1). As a result the very likely
reduction of the theoretically allowed upper bound on the MSSM
lightest Higgs mass $m_h$ down to $M_Z$ is expected [22]. If so, the
Higgs boson $h$ might be accessible at LEPII.

To conclude the most crucial prediction of the presented SU(8) model
surely belongs to the very existence of the PGB's and their
superpartners, just three families of them. Depending on the details
of their mixing pattern with ordinary Higgs sector of MSSM they could
influence appreciably on the particle phenomenology expected at the
TeV scale. We will specially address this interesting question
elsewhere.

\section{Summary}

In this paper we have investigated the supersymmetric SU(N) GUTs
where the missing VEV solution to the doublet-triplet splitting
problem could naturally appear.

In contrast to a conclusion that "the unitary groups with adjoint
breaking do not look very promising in this regard" [4] the missing
VEV configurations actually develop in the two (and more) adjoint
scalar case. We found that the numbers of the fundamental colours and
flavours are happened to be equal in all those vacuum configurations
and for the even-order SU(N) groups (N=2k) the $n_w=2$ configuration
leads to the absolute minimum of potential $V_{adj}$ (9) in the
parameter natural area (12).

Contrary to the supersymmetric adjoint breaking  the SU(N) flavour
subsymmetry breaking, while at a high scale $M_f\sim M_G$,
is triggered just by SUSY breaking. Following such a scenario we have
shown that the flavour part of the superpotential $W_{\varphi}$ (16)
having been constrained by some generic string-inspired extra
$Z_3$-symmetries (accompanied by an accidental $SU(N)\otimes
SU(N)$) does not affect markedly the basic adjoint vacuum
configuration in the model. Another flavour symmetry-breaking
scenario$^4$ using the anomalous $U(1)_A$ also seems well suited to
keep the missing VEV configuration going down to low energies.

In essence, there is only one parameter left for a final
specialization theory, a number of colours $n_c$ (see Eq.(8)), after
which one comes to the distinguished SU(8) symmetry case with the
globally dominant missing VEV breaking channel (12) where the flavour
subsymmetry $SU(3)_f$ subsequently breaks (14). The special assignment
(24a) for the basic superfields (quarks and leptons and their
superpartners) allows to consider the above flavour symmetry as a
chiral SU(3) family symmetry [15] treating families as its own
fundamental triplets (24b).

Thus, according to our starting motivation (Sec.1) the presented
SU(8) model meet a natural conservation of flavour both in the
particle and sparticle sectors, respectively.  Among its direct
predictions the most crucial one belongs to an existence on the TeV
scale just three families of pseudo-Goldstone bosons and their
superpartners (25) which properly improves the
unification picture of MSSM (Table 1). Another significant outcome of
the model turns out to be the very likely reduction of the
theoretically allowed upper bound on the MSSM lightest Higgs mass
$m_h$ down to $M_Z$ what could make $h$ boson to be detectable at
LEPII.

In conclusion, one could hopefully think that the gauge hierarchy
phenomenon in the SUSY SU(N) GUTs gives some insight into the both
salient feature of SM just as an interplay between colours and
flavours (or families for due assignment quarks and leptons) so the
weak SU(2) symmetry structure coming safely from a grand unified
scale down to low energies.

\vspace{1cm}
{\bf Acknowledgements}

\vspace{0.3cm}
One of us (J.L.C.) is deeply grateful to R.Barbieri, S.Ranjbar-Daemi
and A.Smirnov for helpful discussions and stimulating comments and
S.Ranjbar-Daemi also for warm hospitality at ICTP High-Energy
Department where part of this work was carried out. We have greatly
benefited from many interesting discussions with Z.Berezhiani and
I.Gogoladze.

This work was supported in part by the grant of the Georgian Academy
of Sciences and the INTAS grant RFBR 95-567.


\end{document}